# Intrinsic resolving power of XUV diffraction gratings measured with Fizeau interferometry

Samuel Gleason[a], Jonathan Manton[a], Janet Sheung[a], Taylor Byrum[a], Cody Jensen[a], Lingyun Jiang[a], Joseph Dvorak[b], Ignace Jarrige[b], Peter Abbamonte[a,c,*]

[a]Inprentus, Inc., 51 E. Kenyon Rd., Champaign, IL, 61820, USA
[b]NSLS-II, Brookhaven National Laboratory, Upton, NY, 11973, USA
[c]Seitz Materials Research Laboratory, University of Illinois, Urbana, IL, 61801, USA

**ABSTRACT**

We introduce a method for using Fizeau interferometry to measure the intrinsic resolving power of a diffraction grating. This method is more accurate than traditional techniques based on a long-trace profiler (LTP), since it is sensitive to long-distance phase errors not revealed by a *d*-spacing map. We demonstrate 50,400 resolving power for a mechanically ruled XUV grating from Inprentus, Inc.

## 1. INTRODUCTION

Diffraction gratings exhibiting high efficiency in the soft x-ray and ultraviolet (XUV) regime are of rapidly growing importance for next-generation x-ray light sources, such as synchrotron and free-electron laser (FEL) facilities, deep ultraviolet (DUV) and extreme ultraviolet (EUV) semiconductor lithography, and spectroscopic studies of solar weather that influences the Earth's climate, among other subjects. Blazed gratings are preferred over laminar gratings for most applications because they exhibit higher efficiency [1].

Traditionally, diffraction gratings are manufactured using interference lithography, which creates uniform line patterns with near-perfect periodicity. Such holographic approaches generate laminar grooves but may be post-processed with ion milling to create blaze patterns [1]. Unfortunately, gratings for XUV light usually require blaze angles below 1°, which can be prohibitively difficult to achieve using holographic techniques. Further, because refractive optics do not exist in the XUV regime, the modern trend is toward the use of variable-line-spacing (VLS) gratings, in which the periodicity is varied across the surface to achieve focusing as well as wavelength dispersion. While interference lithography can create an essentially perfect uniform pattern, it can only approximate a VLS pattern, creating difficulties for achieving high resolving power.

For this reason, there is renewed interest in fabricating gratings via mechanical ruling, which can create blaze angles below 1° by burnishing. Mechanical ruling can also create an exact VLS pattern, since each line is positioned individually. However, for ruling to be viable for state-of-the-art XUV spectrometers—some of which aim to achieve a resolving power $R \sim 100{,}000$ [2]—each line must be positioned with an accuracy better than 50 nm over areas approaching 20 cm. This creates a need for improved, phase-sensitive metrology techniques capable of quantifying the long-distance coherence properties of grating patterns.

Here, we show that Fizeau interferometry is the method of choice for resolving-power metrology of diffraction gratings. Carried out in Littrow geometry, in which the diffracted beam is back-reflected along the incident beam path, Fizeau interferometry fully quantifies the long-distance phase errors that limit the resolution of a grating pattern. Such nonlocal phase errors are not directly revealed by, for example, long trace profiler (LTP) methods, which are optimized to reveal local properties. A key advantage of the Fizeau approach is that it allows the intrinsic resolution of a grating to be measured without the need for using it in a spectrometer for experiments. Moreover, Fizeau interferometers are modestly priced, commercial instruments that may be operated in a standard laboratory environment without significant investments in time or capital. We demonstrate this technique by applying it to two test grating patterns mechanically ruled by Inprentus, Inc., in 2015, one of which showed a resolving power $R = 50,400$.

---

[*] abbamonte@mrl.illinois.edu

## 2. WHAT DETERMINES THE RESOLVING POWER OF A GRATING?

The purpose of a diffraction grating is to provide wavelength dispersion. Parallel rays of incident light with different wavelengths will diffract from a grating at different angles, allowing polychromatic light to be separated into its spectral components. In a real optical system a grating only works when used in conjunction with optics that focus these diffracted rays onto an image plane, which can be accomplished using lenses, curved mirrors, or the grating itself by using a VLS pattern. In this article, we will be concerned only with the dispersive property of gratings and how the wavelength resolution is compromised by errors in the line positioning. We will therefore focus on uniform gratings, in which the $d$-spacing is (approximately) constant. Our results may be generalized later to VLS patterns in a straight-forward way.

A diffraction grating should be thought of as a one-dimensional crystal. How line positioning errors influence its ultimate resolving power may be understood by analogy with x-ray diffraction from disordered crystals, in which strain results in angular broadening of Bragg reflections [7].

### 2.1 Diffraction from a periodic object

Before discussing imperfections, we review the basics of diffraction of light from a periodic structure. Light diffracts when it encounters an object whose dielectric function is nonuniform in space [3,4]. In the local approximation, the dielectric function has the form

$$\varepsilon(\mathbf{r}) = 1 + \chi(\mathbf{r}), \tag{1}$$

where $\chi(\mathbf{r})$ is the spatially varying dielectric susceptibility of the system. In the far-field limit, the amplitude of the diffracted electric field is proportional to the Fourier transform of the susceptibility,

$$\chi(\mathbf{q}) = \int d\mathbf{r}\, \chi(\mathbf{r})\, e^{-i\mathbf{q}\cdot\mathbf{r}}, \tag{2}$$

where $\mathbf{q} = \mathbf{k}_i - \mathbf{k}_f$ is the momentum transfer, $\mathbf{k}_i$ and $\mathbf{k}_f$ being the wave vectors of the incident and scattered light, respectively [5,6].

In the case of diffraction from a perfectly periodic object the scattering is concentrated into discrete peaks. This can be understood by recognizing that a crystal is composed of many copies of the same object arranged in a lattice, i.e.,

$$\chi(\mathbf{r}) = \sum_n \chi_0(\mathbf{r} - \mathbf{R}_n), \tag{3}$$

where $\chi_0(\mathbf{r})$ represents the susceptibility of a single repeating unit and $\mathbf{R}_n$ are the Bravais vectors of the periodic lattice. The Fourier transform of Eq. 3 is the sum of a set of phases, i.e.,

$$\chi(\mathbf{q}) = \chi_0(\mathbf{q}) S(\mathbf{q}) \tag{4}$$

where

$$S(\mathbf{q}) = \sum_n e^{-i\mathbf{q}\cdot\mathbf{R}_n} \tag{5}$$

is called the structure factor. $S(\mathbf{q})$ is large whenever the momentum transfer, $\mathbf{q}$, is equal to a reciprocal lattice vector, $\mathbf{G}$, and is small otherwise. By definition, $e^{-i\mathbf{G}\cdot\mathbf{R}_n} = 1$ for all $\mathbf{G}$ vectors and $\mathbf{R}_n$, so $S(\mathbf{G}) = N$, where $N$ is the number of repeating units in the crystal. If $\mathbf{q} \neq \mathbf{G}$ then $S(\mathbf{q})$ is small, since the phase factors average to zero. In other words, the diffraction from a periodic object is concentrated into discrete reflections, which for a crystal are called Bragg peaks and for a grating are called diffraction orders. Rays that exhibit $\mathbf{q} = \mathbf{G}$ are said to satisfy the Bragg condition. These reflections are not infinitely narrow, but exhibit a finite width determined by the number of repeating units [7].

The periodic character of a crystal allows the inverse transform of Eq. 4 to be written as a discrete series,

$$\chi(\mathbf{r}) = \chi_0(\mathbf{G}) \sum_{\mathbf{G}} e^{i\mathbf{G}\cdot\mathbf{r}} . \tag{6}$$

In x-ray diffraction from crystals, the Born approximation is usually valid and the integrated intensity of Bragg reflection $\mathbf{G}$ is proportional to the squared amplitude, $|\chi_0(\mathbf{G})|^2$. This is the principle behind x-ray crystallography [7]. In the case of a grating, the Born approximation breaks down and recursive, multiple-scattering calculations are needed to compute the diffracted intensity [4]. However the Bragg condition, $\mathbf{q} = \mathbf{G}$, still applies since it follows from momentum conservation which must hold even in the limit of strong multiple scattering.

In the case of a grating, the periodicity is one-dimensional and the above quantities may be written solely in terms of the coordinate along the surface, $w$, e.g., $\chi = \chi(w)$. The reciprocal lattice vectors are indexed by a single integer, $m$, i.e., $G = 2\pi m/d$, where $d$ is the grating period. Light will reflect when the component of its momentum transfer parallel to the surface corresponds to a reciprocal lattice vector,

$$q = \mathbf{k}_i^{\|} - \mathbf{k}_f^{\|} = \frac{2\pi m}{d} , \tag{7}$$

where $\mathbf{k}_i^{\|} = k\sin\theta_i$ and $\mathbf{k}_f^{\|} = k\sin\theta_f$ are the projections of the initial and final momenta along the surface, and $k = 2\pi/\lambda$ is the wave vector of the light. Eq. 7 is usually written as the grating equation,

$$d\left(\sin\theta_i - \sin\theta_f\right) = m\lambda \tag{8}$$

where $m$ is the diffraction order.

At this point it is evident that no grating can exhibit perfect resolution. A grating provides dispersion because all rays must satisfy the Bragg condition (Eq. 8), requiring rays with different momenta to diffract at different angles. However, a grating reflection exhibits a nonzero, intrinsic width defined by the number of lines in the pattern, $\Delta q = q/(N \cdot m)$. The resolving power of a spectrometer is normally defined as $R \equiv E/\Delta E$. From Eqs. 7 or 8 it is evident that $E/\Delta E = q/\Delta q = N \cdot m$, which for $m = 1$ gives the well-known result that the resolving power of a grating is ultimately limited by the number of lines, even if the grating is perfect [8].

A diffraction grating is usually used only in a single order, $m$. Therefore only one Fourier component of the susceptibility is important for determining its resolution. Hence, for most purposes, we can drop all terms in Eq. 6 except for those associated with the order of interest,

$$\chi(w) = \chi_0(2\pi m/d)\left[e^{i2\pi mw/d} + e^{-i2\pi mw/d}\right] = 2\chi_0(q_m)\cos(q_m w) , \tag{9}$$

where $q_m = 2\pi m/d$ is the wave vector of the diffraction order (note that the effective line density of order $m$ is given by $m/d$).

## 2.2 Diffraction from an aperiodic object

If a grating is imperfect its reflections will be broadened and its resolving power, $R \equiv q/\Delta q$, will be degraded. Whether made by holography or mechanical ruling, the most severe imperfections for the resolving power are errors in the positioning of the lines. We therefore need to generalize the discussion of Section 1.1 to the case in which a grating pattern is slightly aperiodic. Here we will be primarily interested in gratings that are nearly perfect, i.e., in which line positioning errors are small but still are preventing the optic from achieving its ultimate performance.

The first step is to identify the quantity of interest that characterizes positioning imperfections in a grating. Diffraction is a nonlocal, collective effect that depends on the long-distance phase properties of the grating pattern. Local properties, such as a histogram of $d$-spacings generated by a long-trace profiler (LTP), do not contain phase information and are not a direct probe of resolving power. For example, a perfect grating examined under an LTP would exhibit an infinitely narrow histogram of $d$-spacings, which might lead one to conclude that its resolving power is infinite. But the resolving power, $R < N\,m$, is always limited by the number of lines.

The quantity that correctly describes positioning errors in a grating was introduced in the context of phase slips in charge density wave materials by McMillan [9], who argued that periodicity errors are best quantified by a spatially varying phase, $\theta(w)$, which enters Eq. 9 as,

$$\chi(w) = 2\chi_0(q_m)\cos\left[q_m w - m\,\theta(w)\right]. \tag{10}$$

The physical meaning of $\theta(w)$ is the distance, at location $w$, that the line pattern is displaced parallel to the surface from its ideal position, measured in radians. For example, if the pattern at $w$ is in its correct location, then $\theta(w) = 0$. If the pattern at that point is displaced by exactly one period, then $\theta(w) = 2\pi$. As we will see below, the function $\theta(w)$, which we will refer to here as a phase map, fully quantifies the long-distance phase properties that define the resolving power of a grating.

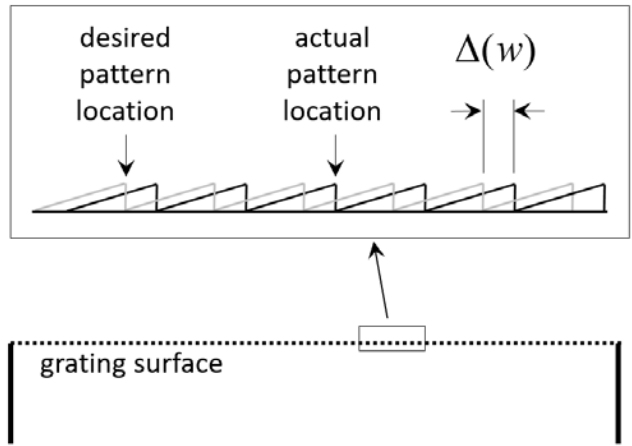

Figure 1. Physical meaning of the displacement map, $\Delta(w)$. Line positioning errors in a grating will, if viewed close-up, look like a perfect pattern (black sawtooth) displaced from its desired location (gray sawtooth). $\Delta(w)$ represents the value of this displacement at location $w$.

It is also useful to examine the normalized quantity,

$$\Delta(w) = \frac{d}{2\pi} \theta(w), \tag{11}$$

which we call a displacement map (see Fig. 1). The quantity $d$ in Eq. 11 should now be taken to represent the mean periodicity of the pattern. Physically, $\Delta(w)$ represents the distance the lines at point $w$ are displaced from their ideal location in the plane parallel to the surface, and quantifies the actual location of the ruling tip during fabrication. It may be used to compare the displacement errors in gratings with differing periods. Note that both $\theta(w)$ and $\Delta(w)$ are definable only on a length scale much larger than $d$, meaning they should be thought of as coarse-grained quantities like temperature or density.

We are interested in how the displacement errors quantified by $\theta(w)$ influence the diffraction lineshape (we discuss how to measure $\theta(w)$ in Section 3). This comes by Fourier transforming Eq. 10, which gives

$$\chi(q) = \chi_0(q_m) S(q), \tag{12}$$

where

$$S(q) = \int dw\, F(w) e^{-iqw} \tag{13}$$

is the structure factor. The quantity in the integrand is called the scattering function,

$$F(w) = e^{-i m \theta(w)}. \tag{14}$$

In obtaining Eqs. 12-14 we have dropped the $-q_m$ order and shifted the momentum origin to measure $q$ with respect to $q_m$. The scattered intensity from the grating is proportional to the square of the **E** field, which in arbitrary units is given by the square modulus

$$I(q) = |S(q)|^2. \tag{15}$$

Eq. 15 is called the resolution function of the grating. It represents the intrinsic spectral lineshape that would limit the energy resolution of a spectrometer in which the grating is installed. Because the scattering from a grating is maximal when $q = q_m$, i.e., when the Bragg condition is satisfied, Eq. 15 will exhibit a peak centered at $q = 0$. The width of this this peak, $\Delta q$, determines the resolving power,

$$R = \frac{q_m}{\Delta q}. \tag{16}$$

Note that $I(q)$ and $R$ do not depend on the wavelength or experimental geometry in which the grating is used, but are intrinsic properties of the pattern itself.

It is important to check Eqs. 13-16 for a known case, so we consider a perfect grating for which $\theta(w) = 0$. The size of the grating pattern $D = 2\pi N / q_m$, where $N$ is the number of lines and $q_m$ is the wave vector of the reflection order of interest. The structure factor for this case evaluates to

$$S(q) = \int_{-D/2}^{D/2} dw\, e^{-iqw} = D\,\text{sinc}\left(\frac{qD}{2}\right), \tag{17}$$

which implies a resolution function

$$I(q) = \text{sinc}^2\left(\frac{qD}{2}\right). \tag{18}$$

The momentum width of Eq. 18 is $\Delta q = 2\pi/D$, which gives a resolving power $R = q_m D/2\pi = m\,N$. So Eqs. 13-16 give the expected result that the ultimate resolution of a perfect grating is limited by the number of lines. For an imperfect grating, $\theta(w)$ may be a nontrivial function and $I(q)$ can have a complicated shape, as we demonstrate below.

## 3. MEASURING $\theta(w)$ WITH A FIZEAU INTERFEROMETER

As argued above, $\theta(w)$ is the quantity of interest for an imperfect diffraction grating. If we can measure this quantity, we can determine its resolving power without the need for installing the grating in a spectrometer. Here, we show that $\theta(w)$ can be measured directly using a Fizeau interferometer with the grating mounted in Littrow geometry.

### 3.1 Fizeau interferometry

Fizeau interferometry is a technique for measuring the contour of an optical surface [10]. In a Fizeau instrument, a parallel beam of monochromatic light with wavelength $\lambda$ uniformly illuminates a surface located under a reference flat (Fig. 2). The primary reflected wave from the surface interferes with the back-reflected reference wave, creating an intensity pattern on a screen given by

$$I(x,y) = I_0 \cos^2 \frac{\varphi(x,y)}{2}, \tag{19}$$

where $\varphi(x,y)$ is the phase difference between the two waves at screen location $(x, y)$. The value of $\varphi(x, y)$ is usually understood as arising from the difference in path length traveled by the two waves at $(x, y)$, which depends on the height of the surface at that point, i.e.,

$$\varphi(x, y) = \frac{4\pi}{\lambda} h(x, y) + \varphi_0, \qquad (20)$$

where $h(x, y)$ is a height function describing the contour of the surface, and the phase offset, $\varphi_0$, accounts for the distance, $L$, which is arbitrary (Fig. 2). Note that the primary wave makes a round-trip, so its optical path length is double the distance between the surface and reference flat. So a $2\pi$ change in the Fizeau phase, $\varphi$, implies a height change of only $\Delta h = \lambda / 2$. Most commercial Fizeau instruments are equipped with a phase-shifting capability that unwraps the interference pattern to reconstruct the phase and height functions.

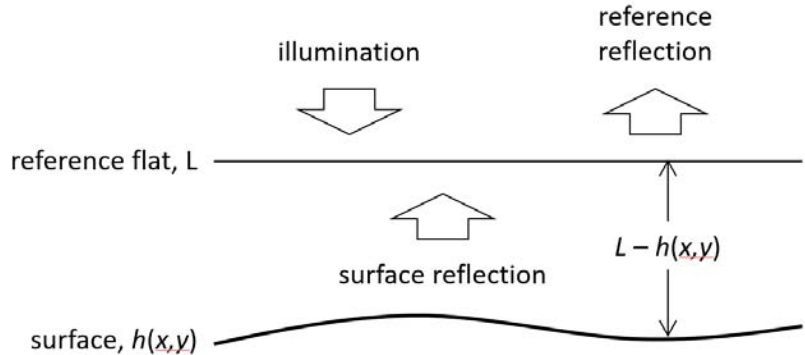

Figure 2. Fizeau interferometry of a surface. The illumination back-reflects from both the surface and the reference flat. The path length difference, $2[L - h(x, y)]$, leads to a phase shift between the two waves, $\varphi(x,y) = 4\pi h(x,y)/\lambda + \varphi_0$, which can be determined from the interference pattern on the screen.

Two subtle points should be understood about Fizeau interferometry. The first is that Eq. 20 is only approximate. Using a full wave mechanics analysis, one can show that Eq. 20 is rigorously correct only in the small-angle limit, so does not apply to surface features that are extremely steep.

Second, the tilt angle in Fizeau interferometry is ambiguous. That is, rotating the optical surface by a small angle creates a linear contribution to $h(x, y)$ that the instrument could interpret either as a tilt or as part of the height function itself. Most Fizeau instruments subtract a plane from $h(x, y)$ so that its average value is zero, nulling out whatever tilt may be present. We will see below that, when using a Fizeau to study gratings, the same effect leads to ambiguity in the average $d$ spacing of the pattern, which must be determined by other means. Fortunately, conclusions reached about the resolving power are not particularly sensitive to the value of $d$.

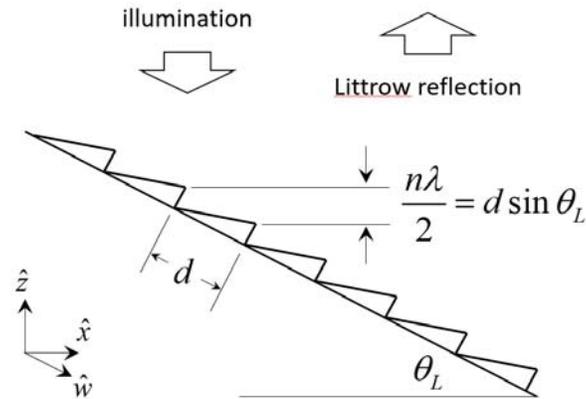

Figure 3. Stroke's staircase picture of a grating in the Littrow configuration [8]. If a grating is tilted to the Littrow angle, the height of each groove along the optic axis, $\hat{z}$, is offset from the previous groove by exactly $n\lambda / 2$, where $n$ is the Littrow order. The back-reflected wave from each groove differs in phase from its neighbors by $2\pi n$. This staircase is therefore indistinguishable from a flat surface in a Fizeau measurement.

## 3.2 Fizeau measurements of a perfect grating in Littrow geometry

It is possible for a diffraction grating to satisfy what is called the Littrow condition, in which the diffracted beam travels back along the same path as the incident. This occurs when the grating is held at the Littrow angle, defined by $\theta_L = \theta_i = -\theta_f$, which from Eq. 7 gives

$$\sin\theta_L = \frac{n\lambda}{2d}, \tag{21}$$

where we use *n* to denote the grating order used in the Littrow measurement, to distinguish it from the order used in experiments, *m*. Eq. 21 indicates that a distinct Littrow condition exists for each order for which $n\lambda/2d < 1$. If a perfect grating is placed under a Fizeau interferometer at exactly the Littrow angle, it will create a back-reflected plane wave indistinguishable from that due to a flat mirror.

As originally pointed out by Stroke [8], this effect can be understood by thinking of the grating as a staircase (Fig. 3). In Stroke's picture, by analogy with the height maps of Section 3.1, each groove of the grating back-scatters the primary wave, the path length difference at each groove being determined by its height projected along the optic axis of the instrument, which we call *z* (Fig. 3). In the Littrow condition, the height of each groove differs from its neighbors by $\Delta z = n\lambda/2$, corresponding to an in optical path length difference $\Delta z = n\lambda$ or a phase difference $\Delta\varphi = 2\pi n$. In other words, even though the surface is at an angle, all the waves depart in-phase along the $\hat{z}$ direction, resulting in an intensity pattern, $I(x, y)$, that is uniform and depends only on the distance between the grating and the reference flat. This Stroke staircase is therefore indistinguishable from a flat surface in a Fizeau measurement.

### 3.3 Fizeau measurements of an imperfect grating

An imperfect grating held at the Littrow angle will deviate from the staircase shown Fig. 3 due to errors in the positions of the lines. These errors, characterized by the displacement map, $\Delta(w)$, will alter the phase shift between the diffracted and reference waves, leading to a nontrivial interference pattern in a Fizeau measurement. Following Stroke's picture [8], the Fizeau phase will be determined by the projection of the line displacements along the *z* axis,

$$\varphi_n(x) = \frac{4\pi}{\lambda} \cdot \Delta(x/\cos\theta_L)\sin\theta_L + \varphi_0 = \frac{2\pi n}{d} \cdot \Delta(x/\cos\theta_L) + \varphi_0, \tag{22}$$

where we have written $\Delta$ as a function of $x/\cos\theta_L$ as a reminder that we are interested in it in terms of the grating coordinate, *w*, which is related to the Fizeau coordinate by $w = x/\cos\theta_L$ (Fig. 3). Note that the phase map, $\varphi_n$, depends on the order of the measurement, *n*. A commercial interferometer may output the phase as a height function,

$$h(x) = \frac{n\lambda}{2d}\Delta(x/\cos\theta_L), \tag{23}$$

which can be understood as the distance, projected along the *z* direction, that the lines at location *x* deviate from an ideal staircase.

Eqs. 22-23 show that a Fizeau measurement of an imperfect grating in Littrow geometry directly measures $\Delta(w)$, which is the quantity that properly encodes the long-distance phase errors that define the resolving power of the pattern. The phase function of the grating can be determined from Eq. 11 or directly from the height map,

$$\theta(w) = \frac{4\pi}{n\lambda} h(w\cos\theta_L) = \frac{2\pi}{d} \frac{h(w\cos\theta_L)}{\sin\theta_L}. \tag{24}$$

Eqs. 22 and 24 imply a direct relationship between the phase function output by the Fizeau and the phase map of the grating

$$\theta(w) = \frac{\varphi_n(w\cos\theta_L) - \varphi_0}{n}, \tag{25}$$

which shows that better sensitivity to displacement errors can be achieved by performing Fizeau measurements in higher order, though this comes at the cost of poorer spatial resolution, since the since of the grating image projected on the Fizeau screen will be reduced. Once $\theta(w)$ is determined, the resolution may be determined using Eqs. 13-16.

Two subtle points should be understood about conducting Fizeau measurements in Littrow geometry. The first is that Eq. 22 is only approximate. Using a full wave mechanics analysis, one can show that Eq. 22 is rigorously correct only in the small-angle limit, i.e., in which the $d$-spacing errors of the grating are sufficiently small that the angle of the diffracted beam does not deviate appreciably from $\theta_L$.

Second, just as the tilt angle is ambiguous in a Fizeau measurement of a mirror, the average $d$-spacing is ambiguous in a Fizeau measurement of a grating. For example, if a grating is rotated slightly from the Littrow condition, a continuous band of interference fringes will appear that the Fizeau could interpret either as a rotation angle or as a $\Delta(w)$ that changes linearly across the surface, implying a $d$-spacing that differs from Eq. 21. A commercial instrument will subtract a plane from Eq. 23, in effect referencing $\Delta(w)$ to some average $d$ spacing that must be determined by other means, such as an absolute measurement of $\theta_L$. Fortunately, the resolution function (Eq. 18) is only linearly sensitive to errors in the value of $d$, e.g., a 0.01% error in $d$ would translate to a 0.01% error in value of $R$.

## 3.4 Fizeau measurements of an imperfect grating that is also not flat

If the grating is written on a substrate that is not perfectly flat, the Fizeau phase will have contributions from both the displacement errors, $\Delta(w)$, and the contour of the surface, $h(x, y)$. This can be a problem when creating test patterns on low-cost substrates, since flatness errors could be interpreted as displacement errors in the line pattern. It would therefore be beneficial, not least from a cost point of view, to separate the effects of $\Delta(w)$ and $h(x, y)$ in cases where both contribute to the Fizeau signal.

In the small-angle approximation, Fizeau interferometry is a linear probe in the sense that the measured phase is directly proportional to the quantity of interest, i.e., Eqs. 20 and 22. As long as this approximation is valid, the phase contributions from displacement errors and height errors simply add,

$$\varphi_n(x, y) = \frac{2\pi n}{d}\Delta\left(x/\cos\theta_L\right) + \frac{4\pi}{\lambda}h(x, y)\cos\theta_L + \varphi_0, \tag{26}$$

where $h(x, y)\cos\theta_L$ is the height function of the surface, which has been tilted by angle $\theta_L$, projected along the $z$ direction.

Eq. 26 exhibits an important symmetry. The first term, which comes from the line position errors, is odd in the grating order, $n$, while the second term is even. This means that displacement map can be isolated by computing the antisymmetric quantity [11],

$$\Delta\left(x/\cos\theta_L\right) = \frac{d}{2\pi n}\frac{\varphi_n(x, y) - \varphi_{-n}(x, y)}{2}. \tag{27}$$

Put more simply, it is not possible to distinguish line positioning errors from flatness errors in a single Fizeau measurement. However, if one performs two Fizeau measurements in opposite orders, $+n$ and $-n$, one can isolate the displacement contribution by taking the difference in the phase maps (Eq. 27). Similarly, one can determine the height map from the symmetric combination,

$$h(x, y) = \frac{\lambda}{4\pi\cos\theta_L}\frac{\varphi_n(x, y) + \varphi_{-n}(x, y) - 2\varphi_0}{2}, \tag{28}$$

though this could be done more easily by just looking at the $n = 0$ order in a conventional surface contour measurement as described in Section 3.1.

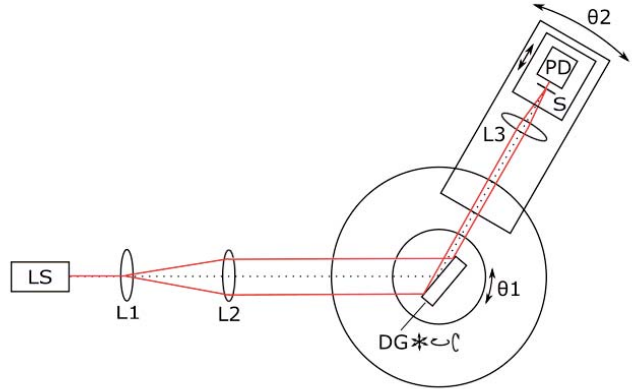

Figure 4. Diffractometer for measuring $I(q)$ (Eq. 15) from the angular distribution of rays diffracted from a grating. A green HeNe laser (LS) is sent through a pair of collimating lenses (L1 and L2) that blow up the beam spatially and narrow its angular spread. The beam strikes the grating (DG), which is mounted on a rotation stage ($\theta_1$) that precisely controls the angle of incidence. The scattered light is measured with a detector assembly comprising a lens (L3) and 5 μm slit (S), which together act as an angle filter, in front of a photodiode (PD). This assembly sits on a second rotation stage that precisely defines the diffracted angle. The overall angular resolution of this setup is about 0.005°, which is adequate for measuring resolving powers up to around 7000 for a 1200/mm grating pattern.

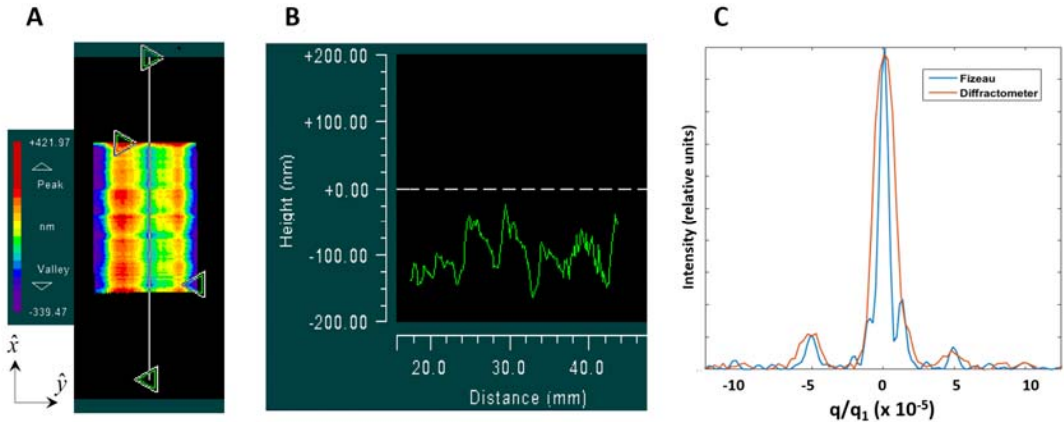

Figure 5. Comparison between Fizeau and diffractometer measurements of the resolution function of a test grating. (A) Raw image showing the $n = 1$ height function output of the Fizeau. The height modulation in the $y$ direction, $\delta\Delta_y$, indicates that the lines are not perfectly straight. This will not affect the resolution of a spectrometer so long as the length of its exit slit in the nondispersive direction is larger than, $W \, \delta\Delta_y / D$, where $W$ is the width of the pattern and $D$ is the distance between the grating and the exit slit. (B) $n = 1$ height map, $h(x)$, obtained by taking a section through the Fizeau image, showing drifts in the line positioning. (C) Resolution functions for $m = 1$ measured with the diffractometer (orange line) and deduced from the Fizeau data (blue line), plotted against the normalized momentum transfer, $q / q_1$. The two curves are in excellent agreement, showing the same asymmetry of the central line, as well as the same satellite structure (the curves were scaled to the same magnitude at $q = 0$). Note that the diffraction data is slightly broadened because of the finite angular resolution of the instrument (Fig. 4).

## 4. EXPERIMENTAL RESULTS

Finally, we demonstrate the Fizeau method by applying it to two mechanically ruled test patters fabricated in 2015 by Inprentus, Inc. Our emphasis will be on validating the Fizeau approach, by comparing it to an independent measure of the resolution function. One such comparison would be to deploy the grating in a spectrometer and perform a measurement of a sharp spectral line whose width is known. But this comparison is not ideal since the resolution function, I(q) (Eq. 15), would be convolved with both the source size and the natural width of the line, which could blur important features in the grating lineshape.

A better approach is to compare $I(q)$ determined from Fizeau to the angular spread of rays diffracted from the grating. If a grating is perfect (and infinite), the grating equation (Eqs. 7 and 8) will be satisfied exactly by all rays. $I(q)$ is a description of the extent to which rays from a real grating deviate from this ideal behavior. Therefore, a direct way to measure the resolution function of a grating is to illuminate it with highly collimated light and precisely measure the angular shape of one of its diffraction orders.

For this purpose, we constructed a visible light diffractometer whose schematic is shown in Fig. 4. The light source is a green HeNe laser (543 nm) collimated to an angular divergence of ~0.003°. This source illuminates the grating whose surface resides at the center of a rotation stage that precisely controls the angle of incidence. The diffracted light is measured with a detector assembly that consists of an angle filter and photodiode, which resides on a second rotation stage that precisely defines the scattering angle. Scanning the detector angle through a grating reflection yields a curve, $I(\theta_2)$, which when plotted against the in-plane momentum transfer gives the resolution function, $I(q)$. This technique is a simple way to perform resolving power metrology on gratings, however its angular resolution (including the detector contribution) is limited to about 0.005°, which allows resolving power measurements only up to about $R = 7000$ for typical gratings. Nevertheless, it provides a means to validate the Fizeau approach outlined in Section 3.

In Figure 5 we compare a Fizeau measurement of the resolving power of a test grating to the lineshape measured with our diffractometer (Fig. 4). The grating was a 5-cm long, 2400/mm, mechanically ruled pattern containing 120,000 lines (Fig. 5a). This pattern was made in 2015 without using interferometer feedback, so it contains line positioning errors that can be quantified as described in Section 3. Fizeau measurements were done on a Zygo Verifire MST phase-shifting interferometer with a 4″ clear aperture. As shown in Fig. 5c, the resolution function, $I(q)$, exhibits an asymmetry as well as two satellite features whose momenta and spectral weight are quantitatively consistent between the two measurements. The only difference between the curves is a visible broadening of the diffraction spectrum compared to the Fizeau curve, which is a consequence of the finite 0.005° angular resolution of our diffractometer.

Three important conclusions should be drawn from Fig. 5. The first is that, apart from finite resolution effects, the two methods for measuring $I(q)$ are quantitatively consistent, validating the Fizeau approach to measuring the resolution function of a grating. Second, Fig. 5c shows that Fizeau provides much better resolution than spectroscopic methods. As a real-space probe, the momentum resolution in a Fizeau measurement is limited only by the field of view, $M$, to be $\Delta q = 2\pi / M$, which can be made arbitrarily large by demagnifying the image. Conversely, the spatial resolution of the Fizeau limits its ability to detect high frequency errors. The highest momentum observable $Q_{max} = \pi / \Delta w$, where $\Delta w$ is the projection of the Fizeau pixel size onto the grating surface. Note that $Q_{max}$, because of the sampling theorem [12], is half the Nyquist frequency, $Q_N = 2\pi / \Delta w$.

Third, and most important, the concept of a resolving power, $R = E / \Delta E$, is too simplistic for many grating patterns. An imperfect grating may not just exhibit a broadened resolution, but can show asymmetry, nontrivial satellite structure, and other effects that cannot be quantified by a single number. Fabricating high-resolution gratings requires understanding these features and engineering a grating lineshape, $I(q)$, that is not only sharp but compact, symmetric, and free of peripheral structure that could create anomalies in high-resolution instruments.

We also demonstrate the antisymmetrization procedure for eliminating flatness errors outlined in Section 3.4. In Figure 6 we show Fizeau measurements of a second grating pattern, this time with a line density of 500/mm. This pattern was 10 cm in size and contained 50,000 lines. We performed Fizeau measurements in both +1 and -1 order and determined $\Delta(w)$ from the difference (Eq. 27; see Fig. 6a,b). The resolution function of this pattern is plotted in Fig. 6c compared to that of a theoretically perfect pattern of the same size. Note that, while the line positioning errors are ~100 nm over the surface of this pattern, its resolution function is nearly perfect, exhibiting the same sinc$^2$-type oscillations expected of an ideal grating (see Eq. 17). The FWHM width of the peak is $\Delta q = 6.23 \times 10^{-8}$ nm$^{-1}$, which translates into a nominal resolving power $R = q_1 / \Delta q = 50{,}400$. The only effect of the line positioning errors is to create a slight asymmetry in the intensity of the finite-size oscillations compared to the ideal curve. This illustrates an important principle in grating fabrication, which is that the resolving power is not just determined by the size of the line positioning errors, but of the statistical distribution of these errors and the type of correlations present.

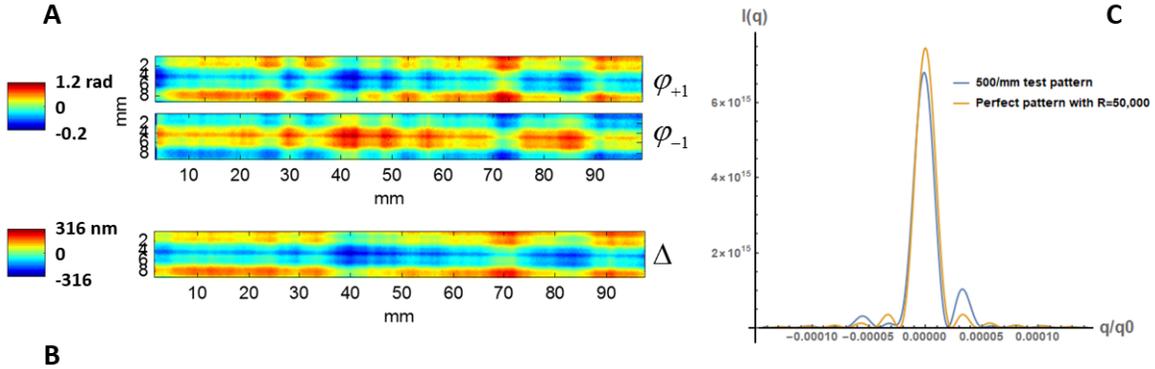

Figure 6. Fizeau measurements of a high resolution, 10 cm, 500/mm pattern containing 50,000 lines. (A) Raw phase output of the Fizeau measurement for both $m = +1$ and $m = -1$ orders. (B) Antisymmetrized displacement map obtained from Eq. 27, which has the contribution from the flatness errors removed. (C) Resulting resolution function, I(q), compared to that for a the theoretical perfect pattern with the same parameters. The FWHM of the peak $\Delta q = 6.23 \times 10^{-8}$ nm$^{-1}$, which translates into a nominal FWHM resolving power $R = q_0 / \Delta q = 50{,}400$.

## 5. CONCLUSIONS

In conclusion, we have introduced a new approach to resolving power metrology for diffraction gratings that uses a Fizeau interferometer. There are two major advantages of this approach. The first is that it is phase-sensitive, properly characterizing the long-distance phase errors that most influence the resolving power and are not (directly) revealed by local probes, such as long trace profilometry. Second, a Fizeau is an inexpensive, off-the-shelf, commercially available instrument, allowing grating metrology to be done without large investments in capital or development time. The approach outlined here applies to the case of uniform gratings, in which the line density is constant over the surface, but it may be generalized to VLS gratings in a straight-forward way.

## ACKNOWLEDGEMENTS

We gratefully acknowledge Z. Hussain, Y. D. Chuang, W. Yang, E. Gullikson, H. Padmore, and D. Voronov for enlightening discussions, and K. Kaznatcheev for sharing the Fizeau data in Fig. 6. This work was supported by National Science Foundation grant SBIR-1248644.